\documentclass[aps,prm,superscriptaddress,reprint]{revtex4-1} 

\usepackage[colorlinks=true,linkcolor=blue,citecolor=blue,urlcolor=blue]{hyperref}

\usepackage{setspace} 
\usepackage{graphicx}
\usepackage{amsmath}
\usepackage{color}
\usepackage{amsmath}
\usepackage{amssymb}
\usepackage{verbatim}
\usepackage{latexsym}
\usepackage{enumerate} 
\usepackage{bm} 
\usepackage{ulem} 
\usepackage[caption=false]{subfig}
\usepackage{multirow}
\usepackage{booktabs}
\usepackage{lineno}

\begin{document}

\title{%
C$_{60}$ building blocks with tuneable structures for tailored functionalities
}

\author{Darius Kayley}
\affiliation{Fitzwilliam College, University of Cambridge, University of Cambridge, Storey's Way, Cambridge CB3 0DG, United Kingdom}

\author{Bo Peng}
\email{bp432@cam.ac.uk}
\affiliation{Theory of Condensed Matter Group, Cavendish Laboratory, University of Cambridge, J.\,J.\,Thomson Avenue, Cambridge CB3 0HE, United Kingdom}

\date{\today}

\begin{abstract}
We show that C$_{60}$ fullerene molecules can serve as promising building blocks in the construction of versatile crystal structures with unique symmetries using first-principles calculations. These phases include quasi-2D layered structures and 3D van der Waals crystals where the molecules adopt varied orientations. The interplay of molecular arrangement and lattice symmetry results in a variety of tuneable crystal structures with distinct properties. Specifically, the electronic structures of these phases vary significantly, offering potential for fine-tuning the band gap for electronics and optoelectronics. Additionally, the optical properties of these materials are strongly influenced by their crystalline symmetry and molecular alignment, providing avenues for tailoring optical responses for photonics. Our findings highlight the potential of fullerene-based building blocks in the rational design of functional materials.
\end{abstract}

\flushbottom

\maketitle

\section{Introduction}

Atoms are the elemental building blocks that govern the chemical bonding and thus the structure of matter. 
The intrinsic properties of the atoms, including the electronic configuration, electronegativity, and bonding tendencies, dictate the preferred packing arrangement and crystal symmetry. As an example, an element like copper naturally adopts a face-centred cubic (fcc) structure and cannot be reconfigured to form a hexagonal close-packed (hcp) lattice\,\cite{Jona2003,Grimvall2012}. This structural constraint significantly limits the design of new functional materials at the atomic level where diverse crystal structures are desired for greater structural and functional tunability. Instead, molecular or colloidal building blocks allow direct control and precise tuning over crystal structures and functionalities\,\cite{Doud2020,Hueckel2021}. However, the complex experimental synthesis and assembly of these building blocks struggle to provide clear answers to the rational design of stable structures with tailored functionalities.

Among all the molecular building blocks, the cage-like, hollow C$_{60}$ fullerene with nearly spherical symmetry\,\cite{Kroto1985} provides an atom-like, stable unit, which can act as a versatile platform for constructing functional materials\,\cite{Kratschmer1990,Stephens1991,Heiney1992,Lamb1993,Nunez-Regueiro1995,Xu1995,Springborg1995,Marques1996,Murga2015}. With inherent symmetry, high stability, and remarkable electronic properties, these fullerene building blocks can form various lattice structures with diverse symmetries, enabling precise control over the physical and chemical properties\,\cite{Hou2022,Peng2022c,Peng2023,Meirzadeh2023,Jones2023,Wu2024a}. This adaptability opens new avenues for material design beyond atoms, offering unprecedented opportunities to explore the interplay between symmetry and functionality.

In this work, we use C$_{60}$ fullerene as an ``ultimate building block'' for designing functional materials. We explore the structural, vibrational, electronic, and optical properties of four phases with distinct crystal symmetries from orthorhombic to trigonal and to cubic. The C$_{60}$ molecules in these structures form various lattices including quasi-2D layered structures and van der Waals plastic crystals with different orientations, leading to distinct lattice dynamics, varied electronic band gaps, and tuneable optical properties. Our work provides promising molecular building blocks for designing tuneable structures with tailored functionalities. 

\begin{figure*}
    \centering
    \includegraphics[width=\linewidth]{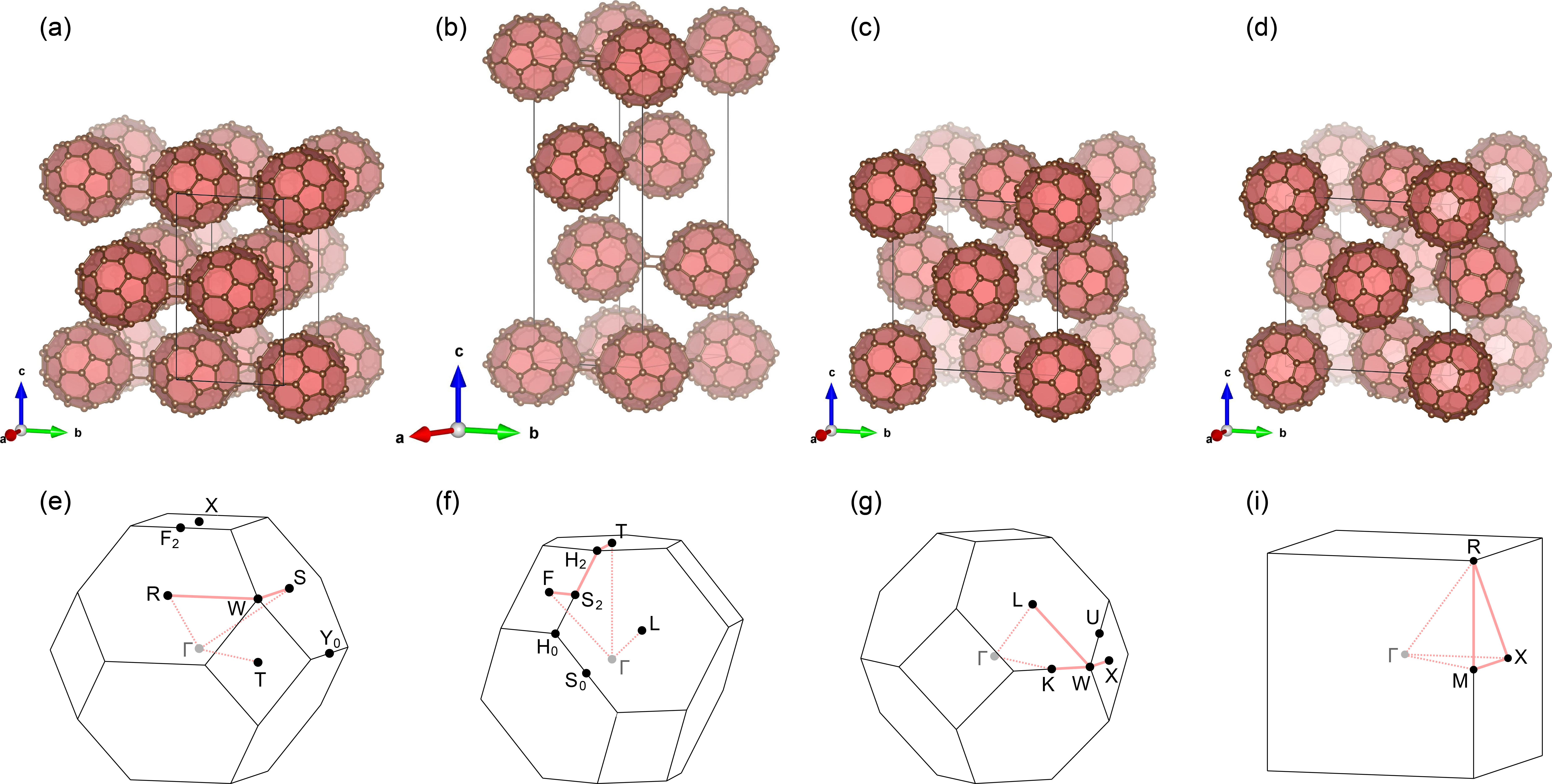}
    \caption{3D and top views of the crystal structures of (a) monolayer and (b) bilayer fullerene networks visualised by {\sc VESTA}\,\cite{vesta}. 
    }
    \label{fig:crystal}
\end{figure*}

\begin{table*}
    \caption{
    Symmetry information, lattice constants with conventional unit cell, volume per C$_{60}$, and cohesive energy $E_{\rm c}$ for different phases of solid C$_{60}$.} 
    \centering
    \begin{tabular}{cccccccc}
    \hline
    Crystal system & Space group & No. & $a$ (\AA) & $b$ (\AA) & $c$ (\AA) & $V$ (\AA$^3$) & $E_{\rm c}$ (eV/atom) \\ 
    \hline
    Orthorhombic & $Immm$ & 71 & 9.038 & 9.137 & 15.289 & 631.309 & -7.66107 \\  
    Trigonal & $R\bar{3}m$ & 166 & 9.190 & 9.190 & 24.796 & 604.489 & -7.65026 \\
    Cubic & $Fm\bar{3}$ & 202 & 14.232 & 14.232 & 14.232 & 720.657 & -7.66258 \\
    Cubic & $Pa\bar{3}$ & 205 & 14.181 & 14.181 & 14.181 & 712.950 & -7.66431 \\
    \hline
    \end{tabular}  
    \label{tbl:crystal} 
\end{table*}

\section{Methods}\label{Methods}

The Vienna \textit{ab initio} Simulation Package ({\sc VASP})\,\cite{Kresse1996,Kresse1996a} was used for first-principles calculations. The projector-augmented wave (PAW) basis set\,\cite{Bloechl1994,Kresse1999} was used with the Perdew-Burke-Ernzerhof (PBE) functional\,\cite{Perdew1996} under the generalized gradient approximation (GGA). The zero damping DFT-D3 method of Grimme was included to describe the van der Waals interactions\,\cite{Grimme2010}. A plane-wave cutoff energy of 800\,eV was used with the Brillouin zone sampled using a well-converged Monkhorst-pack \textbf{k}-point grid of $3\times3\times3$. Both the lattice parameters and the atomic positions were fully relaxed until the difference between the electronic and ionic steps reached $10^{-6}$\,eV and $10^{-2}$\,eV/\AA\, respectively. 
The vibrational properties were calculated under the harmonic approximation using {\sc PHONOPY}\,\cite{Togo2008,Togo2015} with interatomic force constants computed from a finite difference method\,\cite{LePage2002}. A supercell size of $2\times2\times2$ was used for all three phases with one C$_{60}$ molecule in the primitive unit cell, while a $1\times1\times1$ cell of ($14.181$\,\AA)$^3$ was used for the $Pa\bar{3}$ phase with four C$_{60}$ molecules in the primitive unit cell. Unscreened hybrid functional PBE0\,\cite{Adamo1999} calculations were conducted for electronic structures. This approach has predicted comparable band gaps for monolayer fullerene networks\,\cite{Peng2022c,Jones2023} to those from computationally-heavy many-body perturbation theory\,\cite{Champagne2024}. Excitonic effects were included using the time-dependent Hartree-Fock (TDHF) method based on the Casida equation\,\cite{Sander2017} from a basis of 16 highest valence bands and 16 lowest conduction bands with a $\mathbf{k}$-mesh of $8\times8\times8$ ($4\times4\times4$) for the phases with one (four) C$_{60}$ molecule(s) in the primitive unit cell. The PBE0\,+\,TDHF approach replaces the screened exchange $W$ in the Bethe-Salpeter equation (BSE) with the exchange-correlation kernel to describe the screening of the Coulomb potential. This has yielded exciton binding energies for C$_{60}$ systems\,\cite{Peng2022c,Jones2023} quantitatively agreeable with $GW$\,+\,BSE results\,\cite{Champagne2024} but, crucially, is computationally more feasible.



\section{Results and Discussion}\label{Results}

\subsection{Crystal structures}

Figure\,\ref{fig:crystal} shows the crystal structures of solid C$_{60}$ in different phases. There are two layered structures. The orthorhombic phase in Fig.\,\ref{fig:crystal}(a) crystallises in space group $Immm$ (No.\,71) with a van der Waals layered structure. Within each layer, the C$_{60}$ cages along $a$ and $b$ directions are connected by [$2+2$] cycloaddition bonds, forming a 2D quasi-square lattice. Along the $c$ direction, the layers closely stack in a space-efficient manner. The trigonal phase in Fig.\,\ref{fig:crystal}(b) also has a van der Waals layered structure with space group $R\bar{3}m$ (No.\,166). In the 2D plane, the C$_{60}$ cages form a quasi-triangular lattice linked through [$2+2$] cycloaddition bonds with slightly larger lattice constants $a$ and $b$ (Table\,\ref{tbl:crystal}). The $R\bar{3}m$ phase is in a closely-packed rhombohedral lattice with the smallest volume among all four phases, similar to an hcp lattice. However, it has the highest cohesive energy. These two layered structures can be obtained experimentally in the pressure range of $2-8$\,GPa at $500-900$\,K as mixtures, with higher pressures yielding more $R\bar{3}m$ crystals\,\cite{Iwasa1994,Nunez-Regueiro1995,Murga2015}.

\begin{figure*}
    \centering
    \includegraphics[width=\linewidth]{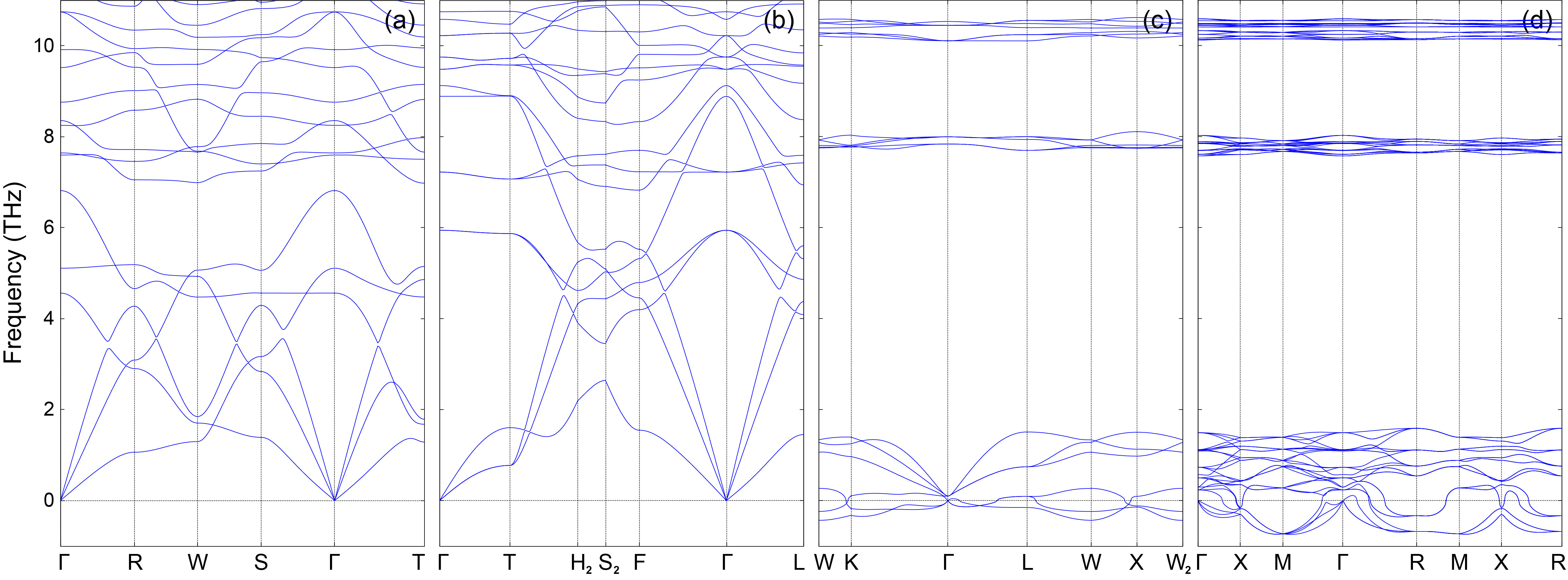}
    \caption{Low-frequency phonon spectra for solid C$_{60}$ with space groups (a) $Immm$, (b) $R\bar{3}m$, (c) $Fm\bar{3}$, and (d) $Pa\bar{3}$.
    }
    \label{fig:phonon}
\end{figure*}

Beyond the layered structures, there are two fcc phases in Fig.\,\ref{fig:crystal}(c) and (d) with no intermolecular bonds. The molecules in these structures, which have much higher volume, are held together purely by van der Waals interactions, with lower cohesive energies than the two layered phases, as shown in Table\,\ref{tbl:crystal}. Considering an isolated fullerene molecule, the icosahedral symmetry leads to a point group of $I_{\rm h}$ ($m\bar{3}\bar{5}$), which, after including translational symmetry that is incompatible with five-fold rotational symmetry, has a maximal subgroup of $m\bar{3}$ crystalline symmetry. The $Fm\bar{3}$ (No.\,202) phase in Fig.\,\ref{fig:crystal}(c) represents the simplest crystalline structure for which all four icosahedra have the same orientation. For the $Pa\bar{3}$ (No.\,205) phase in Fig.\,\ref{fig:crystal}(d), the four sites are inequivalent with different orientations, further reducing the symmetry. As a result of the extra rotational degree of freedom in the C$_{60}$ molecules, the $Fm\bar{3}$ phase is more symmetric than the $Pa\bar{3}$ phase, with one C$_{60}$ molecule in the primitive unit cell, leading to a typical fcc Brillouin zone (BZ), as shown in Fig.\,\ref{fig:crystal}(g). On the other hand, the $Pa\bar{3}$ phase has four inequivalent C$_{60}$ molecules in the primitive unit cell with a cubic BZ, as shown in Fig.\,\ref{fig:crystal}(i). The $Fm\bar{3}$ phase has a slightly larger lattice constant and a higher cohesive energy than the $Pa\bar{3}$ phase.

\subsection{Lattice dynamics}\label{Phonons}

We next examine dynamic stability of the four phases by comparing the phonon spectra. For the two layered structures in Fig.\,\ref{fig:phonon}(a) and (b), the phonons exhibit no imaginary modes, indicating that both structures are in a local minimum of the potential energy surface\,\cite{Peng2017a,Malyi2019,Luo2022,Pallikara2022}. In contrast, both the fcc van der Waals phases are dynamically unstable with imaginary modes across the entire BZ. 

For the $Fm\bar{3}$ phase in Fig.\,\ref{fig:phonon}(c), there are six acoustic-like modes, and the three imaginary modes correspond to the rotational motion of molecules. The rotational disorder corresponds to two possible orientations, leading to an ``average'' molecule with 120 half-occupied atom sites, thus raising the symmetry to $Fm\bar{3}m$\,\cite{Heiney1992}. The orientational disorder can be further increased by assuming completely random orientations, which can be viewed as a spherical shell with the same $Fm\bar{3}m$ symmetry. This crystal symmetry can be found in superconducting K$_3$C$_{60}$\,\cite{Stephens1991,Mitrano2016}. For the $Pa\bar{3}$ phases in Fig.\,\ref{fig:phonon}(d), there are twenty four low-frequency modes as there are four C$_{60}$ molecules in the primitive unit cell, leading to even more complex rotational modes with imaginary frequencies. Experimentally, a phase transition to the $Pa\bar{3}$ phase at 255\,K has been reported\,\cite{Heiney1992}. Therefore, it is expected that the imaginary phonons in this orientationally-ordered phase will disappear after including anharmonic effects at increased temperatures.

\begin{figure*}
    \centering
    \includegraphics[width=\linewidth]{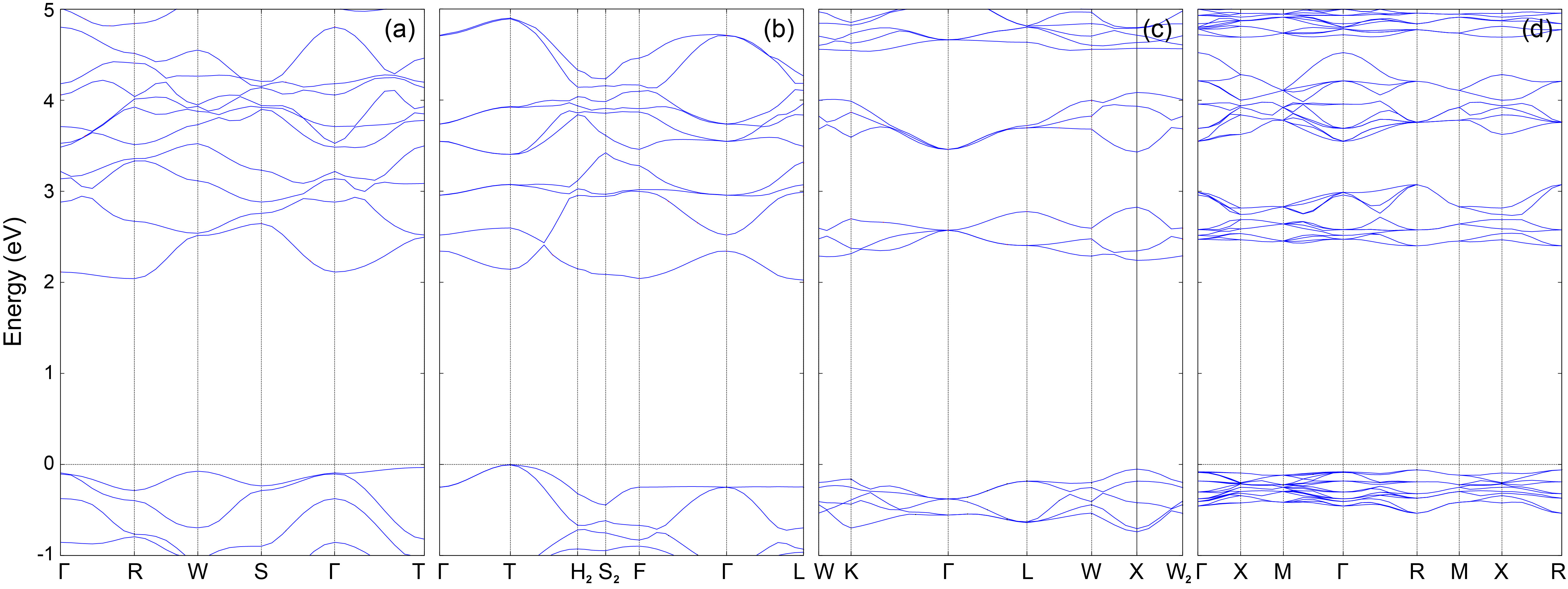}
    \caption{Electronic structures for solid C$_{60}$ with space group (a) $Immm$, (b) $R\bar{3}m$, (c) $Fm\bar{3}$, and (d) $Pa\bar{3}$. 
    }
    \label{fig:electron}
\end{figure*}

\begin{figure*}
    \centering
\includegraphics[width=\textwidth]{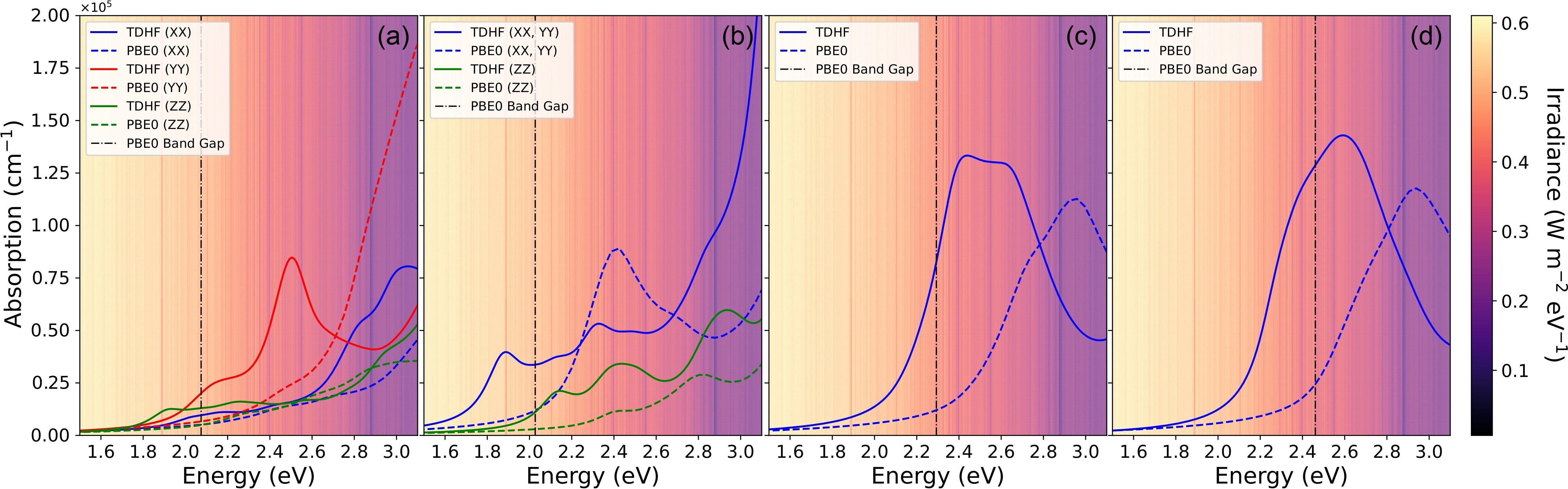}
    \caption{Optical absorption of solid C$_{60}$ with space group (a) $Immm$, (b) $R\bar{3}m$, (c) $Fm\bar{3}$, and (d) $Pa\bar{3}$ without (denoted as PBE0) and with (denoted as TDHF) the excitonic effects. The extraterrestrial solar irradiance \cite{Gueymard2002} is also shown as the background colour map.}
    \label{fig:optic}
\end{figure*}

\subsection{Electronic structures}\label{electronic}

We then compare the electronic structures of the four phases. As shown in Fig.\,\ref{fig:electron}(a), the $Immm$ fullerite has an indirect band gap of 2.075\,eV. The valence band maximum (VBM) is located at the T high-symmetry point, and the conduction band minimum (CBM) is located at the R high-symmetry point [the exact positions of these high-symmetry points are shown in Fig.\,\ref{fig:crystal}(e)]. The $R\bar{3}m$ fullerite in Fig.\,\ref{fig:electron}(b) has an indirect band gap of 2.027\,eV with the VBM at T and the CBM at L. The PBE0 band gaps of the two layered structures are comparable to each other. 

In the two fcc van der Waals structures, the band gaps are much larger than those of the layered structures because of the weaker screening effects with increasing intermolecular distance. The $Fm\bar{3}$ fullerite in Fig.\,\ref{fig:electron}(c) has a direct band gap of 2.293\,eV at X and an energy gap of 2.588\,eV at L, agreeing well with previous $GW$ calculations and  photoemission spectra\,\cite{Shirley1993,Schwedhelm1998,Makarova2001}. The $Pa\bar{3}$ fullerite in Fig.\,\ref{fig:electron}(d) has a direct band gap of 2.461\,eV at R, agreeing well with previous calculations\,\cite{Shirley1996}. 

\begin{table}
    \caption{
    PBE and PBE0 band gaps (eV) for different phases of solid C$_{60}$.} 
    \centering
    \begin{tabular}{ccc}
    \hline
    Space group & PBE $E_{\rm g}$ & PBE0 $E_{\rm g}$ \\ 
    \hline
    $Immm$      & 0.845 (indirect) & 2.075 (indirect) \\  
    $R\bar{3}m$ & 0.742 (indirect) & 2.027 (indirect) \\
    $Fm\bar{3}$ & 1.116  (direct)  & 2.293  (direct)  \\
    $Pa\bar{3}$ & 1.273  (direct)  & 2.461  (direct)  \\
    \hline
    \end{tabular}  
    \label{tbl:band} 
\end{table}


\subsection{Optical Absorption}

Figure\,\ref{fig:optic} summarises the optical absorption of all four phases in the independent-parcticle approximation and in the presence of excitonic effects. For the $Immm$ phase in Fig.\,\ref{fig:optic}(a), the inclusion of excitons leads to a strong optical absorption peak around 1.9\,eV, below the band gap for polarisation along $z$, agreeing well with previous photoemission studies\,\cite{Weaver1991}. The peaks for polarisation along the in-plane directions are mainly above the band gap, and have stronger absorption. For the $R\bar{3}m$ phase, the in-plane polarised light has a strong absorption peak slightly below 1.9\,eV, as shown in Fig.\,\ref{fig:optic}(b). For the van der Waals fcc $Fm\bar{3}$ and $Pa\bar{3}$ phases, the strongest absorption peaks are above the band gap, almost reaching $1.5\times10^5$\,cm$^{-1}$. 

\begin{table}
    \caption{
    Eigenvalue $E_{\rm x}$ (eV) and binding energy $E_{\rm b}$ (eV) of the first bright and first dark excitons for different phases of solid C$_{60}$.} 
    \centering
    \begin{tabular}{ccccc}
    \hline
    Space group & $E_{\rm x}^{\rm bright}$ & $E_{\rm b}^{\rm bright}$ & $E_{\rm x}^{\rm dark}$ & $E_{\rm b}^{\rm dark}$ \\ 
    \hline
    $Immm$      & 1.912 & 0.293 & 1.808 & 0.374 \\  
    $R\bar{3}m$ & 1.871 & 0.275 & 1.900 & 0.347 \\
    $Fm\bar{3}$ & 2.158 & 0.206 & 1.987 & 0.304 \\
    $Pa\bar{3}$ & 2.223 & 0.295 & 2.050 & 0.413 \\
    \hline
    \end{tabular}  
    \label{tbl:exciton} 
\end{table}

We then compare the low-energy excitons by listing the first bright and first dark states. As shown in Table\,\ref{tbl:exciton}, all phases possess a high exciton binding energy above 0.206\,eV, which is expected in molecular systems with weaker screening effects and thus stronger Coulomb interactions between electron-hole pairs. The $R\bar{3}m$ phase has the lowest bright exciton eigenvalue of 1.871\,eV, which is even lower than that of the first dark excitons. This leads to strong exciton absorption below the band gap. In contrast to the $R\bar{3}m$ fullerite, the eigenvalue of the first dark exciton in the other three phases is lower than that of the first bright exciton.

\subsection{Discussion}

The diverse fullerite structural phases with distinct symmetries unlock various applications. One notable application lies in tuning the electronic and optical band gaps, which is critical for optimising performance in semiconductors and photovoltaic devices. By manipulating the crystalline symmetry and the interfullerene distance, electronic properties such as effective masses at band edges can be tuned to enhance carrier mobility and energy conversion efficiency. Additionally, tailored optical properties can be obtained by varying the space groups that govern the optical selection rules. As an example, for the two layered systems with anisotropic optical properties, the orthorhombic $Immm$ phase has stronger out-of-plane exciton absorption, whereas the rhombohedral $R\bar{3}m$ structure exhibits enhanced in-plane absorption in the entire visible spectrum. This enables their applications in the design of polarisation filters and in the directional control of light emission and photodetection. These molecular crystals also offer a promising platform for energy storage such as batteries and supercapacitors because of their large surface area, high chemical stability, and environmental friendliness. The ability to design versatile C$_{60}$ lattices opens a new route towards the development of functional materials for the future, underscoring their transformative potential.

\section{Conclusions}
In conclusion, we demonstrate the versatility of fullerene building blocks for engineering materials with a diverse range of crystal symmetries from quasi-2D layered arrangements to 3D van der Waals plastic crystals. The chemical bonding, intermolecular distance, and orientations of C$_{60}$ molecules play a crucial role in determining the lattice dynamics, electronic structures, and optical properties of these materials. By leveraging the structural tuneability of these phases, we open pathways for designing materials tailored to specific applications in electronics, photonics, and energy systems. Our results show the potential of fullerene-based building blocks as a highly tuneable platform for exploring new physical and chemical phenomena, and developing advanced functional materials. 


\section*{Conflicts of interest}
There are no conflicts to declare. 



\section*{Acknowledgements}
B.P. acknowledges support from Magdalene College Cambridge for a Nevile Research Fellowship. The calculations were performed using resources provided by the Cambridge Service for Data Driven Discovery (CSD3) operated by the University of Cambridge Research Computing Service (\url{www.csd3.cam.ac.uk}), provided by Dell EMC and Intel using Tier-2 funding from the Engineering and Physical Sciences Research Council (capital grant EP/T022159/1), and DiRAC funding from the Science and Technology Facilities Council (\url{http://www.dirac.ac.uk} ), as well as with computational support from the UK Materials and Molecular Modelling Hub, which is partially funded by EPSRC (EP/T022213/1, EP/W032260/1 and EP/P020194/1), for which access was obtained via the UKCP consortium and funded by EPSRC grant ref EP/P022561/1.

\end{document}